**Title:** 2D Semiconductor Nonlinear Plasmonic Modulators

**Author Names:** Matthew Klein[1,2], Bekele H. Badada[1], Rolf Binder[1,2] Adam Alfrey[1], Max McKie[1], Michael R. Koehler[3], David G. Mandrus[3-5], Takashi Taniguchi[6], Kenji Watanabe[6], Brian J. LeRoy[1], and John R. Schaibley[1]

**Author Addresses:**

[1]Department of Physics, University of Arizona, Tucson, Arizona 85721, USA

[2]College of Optical Sciences, University of Arizona, Tucson, Arizona 85721, USA

[3]Department of Materials Science and Engineering, University of Tennessee, Knoxville, Tennessee, 37996, USA

[4]Materials Science and Technology Division, Oak Ridge National Laboratory, Oak Ridge, Tennessee, 37831, USA

[5]Department of Physics and Astronomy, University of Tennessee, Knoxville, Tennessee, 37996, USA

[6]National Institute for Materials Science, Tsukuba, Ibaraki 305- 0044, Japan

**Corresponding Author:** John Schaibley, johnschaibley@email.arizona.edu

**Abstract: A plasmonic modulator is a device that controls the amplitude or phase of propagating plasmons. In a pure plasmonic modulator, the presence or absence of a "pump" plasmonic wave controls the amplitude of a "probe" plasmonic wave through a channel. This control has to be mediated by an interaction between disparate plasmonic waves, typically requiring the integration of a nonlinear material. In this work, we demonstrate the first 2D semiconductor nonlinear plasmonic modulator based on a $WSe_2$ monolayer integrated on top of a lithographically defined metallic waveguide. We utilize the strong coupling between the surface plasmon polaritons (SPPs) and excitons in the $WSe_2$ to give a 73 % change in transmission through the device. We demonstrate control of the propagating SPPs using both optical and SPP pumps, realizing the first demonstration of a 2D semiconductor nonlinear plasmonic modulator, with a modulation depth of 4.1 %, and an ultralow switching energy estimated to be 40 aJ.**

**Main Text:**

**Introduction:**

Plasmonic modulators have been highly sought after for approaches to optical frequency information processing devices[1-5]. Optical frequency plasmonic devices offer potential advantages over electronic devices due to the high carrier frequency of optical waves, as well as the potential to use ultra-fast solid-state nonlinearities for sub-picosecond switching times. Furthermore, by using plasmonic structures, optical frequency waves can be confined to sub-free-space wavelength waveguides allowing for miniaturization of on-chip optical devices[1]. In this letter, we demonstrate a 2D material plasmonic modulator based on controlling surface plasmon polaritons (SPPs) that propagate through metallic waveguides as depicted in Fig. 1a. In order to achieve plasmonic modulation, we integrate a hexagonal boron nitride (hBN) encapsulated semiconductor monolayer transition metal dichalcogenide (TMD) $WSe_2$ on top of the waveguide, where the interaction between SPPs and excitons in the $WSe_2$ provides the nonlinear response needed for modulation.

In free space optical measurements, monolayer $WSe_2$ and other semiconducting TMDs are known to exhibit large light-matter interactions and large third-order nonlinear optical susceptibilities near their exciton resonance[6-12]. Recently, there has been significant interest in using monolayer TMDs for plasmonic applications including the demonstration of SPPs coupling to dark excitons in monolayer $WSe_2$[13,14], increasing the nonlinear response using localized plasmonic effects[15-17], and enhancement of single quantum emitter emission rates[18,19]. In this work, we probe both the linear and nonlinear response of SPPs propagating through hybrid 2D material gold waveguide structures and show that the interactions between SPPs and excitons are surprisingly large. In nonlinear plasmonic measurements, we demonstrate modulation of the SPP amplitude that can be controlled by the intensity and wavelength of either an optical pump or an SPP pump.

Our results rely on the atomically-thin nature of the TMD and the surface confined SPP mode to realize an attractive geometry where the active layer is near the maximum of the SPP mode. Furthermore, we develop a novel self-consistent theory of exciton-SPP (E-SPP) coupling that is unique to the 2D layer geometry and includes a complete E-SPP dispersion relation. We show that our E-SPP model is highly predictive for both the linear (transmission) and nonlinear (differential transmission) response. Unlike previous plasmonic modulators which relied on quantum dots[3], photochromic molecules[20] or high intensity optical pumping of plasmonic waveguides[5], we take advantage of the fast nonlinear optical response of 2D semiconductor excitons to realize an ultra-low switching energy plasmonic modulator with a modulation depth of at least 4.1 %, limited by the pump power used in the measurement.

**Results:**

**Fabrication of E-SPP device and linear response**

Monolayer $WSe_2$ was integrated on top of the metallic waveguide structures to serve as a nonlinear active layer. Monolayer $WSe_2$ was isolated through mechanical exfoliation from high quality bulk crystals. The $WSe_2$ thickness was confirmed by photoluminescence. In order to electrically isolate the $WSe_2$ from the metallic waveguide (to avoid quenching of excitons), it was encapsulated with thin (5 nm top, 2 nm bottom) hBN. The hBN-$WSe_2$-hBN heterostructure was fabricated and transferred onto the waveguide using a polymer based dry transfer technique (polycarbonate film on polydimethylsiloxane, PDMS, stamp)[21]. The transfer was performed under a microscope based probe station to allow for alignment of the 2D heterostructure and waveguide. An optical image of the resulting hybrid 2D material plasmonic structure is shown in Fig. 1b.

The hybrid hBN-WSe$_2$-hBN/plasmonic structures were measured at 4.5 K in a closed-cycle optical cryostat to reduce thermal broadening and achieve the largest possible modulation effect. The transmission spectra were measured using a tunable Ti:sapphire continuous wave laser (M Squared SolsTis). The laser was focused to a diffraction limited spot on the input grating coupler. Light scattered from the output grating coupler was isolated using a spatial filter and measured with a silicon photodiode. In the linear transmission measurements, the probe laser was modulated for lock-in detection. In the nonlinear spectroscopy measurements, pump and probe beams (from two tunable Ti:sapphire lasers) were amplitude modulated at different frequencies near 500 kHz to allow for lock-in detection at the modulation difference frequency.

The SPP transmission spectrum is shown in Fig. 1c for 60 µW input power. The black data show the transmission spectrum for the hybrid hBN-WSe$_2$-hBN/plasmonic structure (with a ~4 µm long WSe$_2$ layer), and the red data show a reference bare waveguide. At the exciton resonance (1.737 eV, 713.6 nm), the transmission is reduced by approximately 73 % due to the presence of the WSe$_2$ layer, indicating a large interaction between SPPs and WSe$_2$ excitons. By comparing these transmission data to the photoluminescence spectrum (Fig. 1c inset), we can identify the dip in the SPP transmission as originating from the WSe$_2$ neutral exciton (X$^0$)[7]. We note that the center energy of the PL and SPP absorption response are aligned to within 1 meV, consistent with previous optical measurements on monolayer WSe$_2$[7].

**Theory of E-SPP**

In order to understand the coupling between SPPs and excitons, we use an extension of the well-known SPP dispersion ($k_x(\omega)$) that relates the wave vector component $k_x$, where the axis $x$ is shown in Fig. 1a, of a mode propagating along the surface to its energy ($\hbar\omega$) where $\omega$ is the SPP's angular frequency. Our approach complements other theories such as coupled oscillator (plexciton)[22,23], scattering[24], and gain-assisted SPP theories[4,25,26]. Using the dielectric function of the metal $\varepsilon_m(\omega)$ and the optical susceptibility $\chi(\omega)$ of the TMD layer, we obtain their coupling directly by solving the dispersion relation, which is free of any fitting parameters. We use subscripts 1, 2, 3 to denote the region above the TMD layer, between the metal surface and the TMD layer, and inside the metal, respectively (Supplementary Fig. 1).

The dispersion relation of the coupled exciton surface plasmon polariton is obtained in a way that is analogous to deriving that of an SPP, i.e. looking for non-trivial solutions of Maxwell's equations that satisfy the continuity relations at the surface and decay away from it. The difference is the presence of the TMD layer, which requires additional continuity relations to be fulfilled, and there is no exponential decay in the region between the metal surface and the TMD layer. In this way, we obtain $k_{2z}(\varepsilon_m k_{2z} - k_{3z}) + g k_{3z} = 0$ where $k_{iz} = \pm\sqrt{\omega^2 \varepsilon_i / c^2 - k_x^2}$ is the wave vector component perpendicular to the surface, the dielectric functions are $\varepsilon_1 = \varepsilon_2 = 1$ for the vacuum regions and $\varepsilon_3 = \varepsilon_m$, and $g = 4\pi i \left(\omega^2 / c^2 - k_x^2\right) \chi(\omega)$ provides the coupling. Here, we assume the distance between the TMD layer and the metal surface to be negligible. A more general form where the distance is an arbitrary input parameter is given in the Supplementary Section 1. We use a Drude model for the metal and a Lorentz model for the TMD exciton. This results in an E-SPP resonance at the exciton energy (1.737 eV) with the real and imaginary parts of $k_x(\omega)$ shown in (grey and blue curves, Fig. 2a.). The E-SPP group velocity is plotted in Supplementary Fig. 2. The peak value for $\mathrm{Im}\, k_x$ of 0.07 µm$^{-1}$ corresponds to an absorption length of 7.1 µm. Fig. 2b shows the measured transmission as a function of the effective WSe$_2$ sample length for the three different structures we investigated (Supplementary Fig. 3a-d). The effective WSe$_2$ sample lengths were estimated from the optical microscope images by calculating an average length over the central 3 µm of the waveguide corresponding the full-width half-max of the SPP spatial mode. An exponential fit to these data yields an effective decay length of 4.2 ± 0.6 µm, which is within a factor of two of our theoretical model.

**Nonlinear E-SPP interactions**

The transmission of the plasmonic device can be controlled by optically pumping the $WSe_2$ excitons, partially saturating the absorption. Fig. 3a depicts the experimental configuration where SPPs propagating through the plasmonic structure serve as the probe, and a free space laser focused on the hBN-$WSe_2$-hBN structure serves as an optical pump. Here, the focused optical pump beam diameter was chosen so that it illuminated nearly the entire $WSe_2$ region with an intensity of 8.5 x $10^6$ W/m². Fig. 3b shows the DT/T spectrum as a function of probe wavelength, i.e., the pump induced differential transmission DT normalized by the probe transmission (T). DT/T spectra for three different pump energies of 1.717 eV (red), 1.739 eV (black) and 1.746 (blue) are shown. When the pump laser is near resonance with the $WSe_2$ $X^0$, the DT/T signal is maximized giving a peak value of DT/T = 4.1 x $10^{-3}$. Fig. 3c shows the pump power dependence of the DT signal near the center of the exciton peak (1.739 eV pump, 1.743 nm probe). The DT signal is linear with pump power indicating that the DT response arises from the third-order nonlinear susceptibility.

In order to demonstrate plasmonic modulation, we performed nonlinear measurements where both pump and probe lasers were coupled into the input grating, launching pump and probe SPPs (depicted in Fig. 4a). Fig. 4b shows the DT/T spectra for three different pump SPP energies 1.715 eV (red), 1.739 eV (black), and 1.771 eV (blue). We again observe a strong nonlinear response at the $X^0$ resonance, corresponding to a maximum DT/T = 4.1 x $10^{-2}$. For this case, we see the DT/T amplitude increases by a factor of 10 over the optically pumped signal. The SPP pump power dependence is also linear (Fig. 4c), consistent with a third-order nonlinear response. From the finite difference time domain (FDTD) model, we find that the SPP pump intensity is 4.6 x 10$^6$ W/m$^2$, ~2 times smaller than the optical pump case. We note that in SPP pump-SPP probe measurements both pump and probe lasers were detected simultaneously, which both contribute to the DT signal. To account for this, the transmission used to calculate DT/T is the sum of both pump and probe beams combined.

To understand this plasmonic modulation effect and to estimate the order-of-magnitude of the third-order nonlinearity, we extended our linear analysis (Fig. 2a) to the third-order nonlinear response with perturbation theory. Since we only observe significant signal near 1.737 eV, we limit our model to in-plane dipoles associated with the X$^0$ excitons. This reduces the susceptibility tensor to the single third-order component $\chi^{(3)}$ (Supplementary Section 1). We assume that the pump-induced change in the susceptibility ($\Delta\chi$) is proportional to the average pump intensity ($I_p$), i.e. $\Delta\chi(\omega, I_p) \approx \frac{4\pi}{c} I_p \chi^{(3)}(\omega)$. We can then use the linear dispersion relation with the replacement $\chi(\omega) \to \chi(\omega, I_p) = \chi(\omega) + \Delta\chi(\omega, I_p)$ which yields a pump-induced change in the dispersion, $\Delta k_x$, and thus a measure of the pump-induced differential transmission DT/T. We deduce a value for Im $\chi^{(3)}$ at the peak of the E-SPP resonance by using the experimental value for DT/T and the estimated average intensity. For both, optical pump/SPP probe and SPP pump/SPP probe we find the order of magnitude of Im $\chi^{(3)}$ to be $-10^{-20} \frac{\text{m}^3}{\text{V}^2}$, in agreement with previously reported all-optical experiments[9,27].

**Discussion:**

In this work, we have investigated both the linear and nonlinear response of excitons interacting with propagating SPPs in metallic waveguide structures. We show that the linear absorption of SPPs can be very large, exceeding 73 %. The large absorption and nonlinear response might be surprising considering that the out-of-plane spatial extent of an SPP (~500 nm) is much larger than the TMD thickness (0.7 nm). However, our theoretical analysis is consistent with the measurements yielding an absorption coefficient on the order of 0.2 µm$^{-1}$. The key to the large linear absorption is the nanometer-scale proximity of the TMD layer to the metal surface, which allows for the active layer to be located near the maximum of the SPP mode. By performing both optical pump and SPP pump DT measurements, we demonstrate control of SPP propagation with a large DT/T response exceeding 4 %. The modulation depth achieved in our monolayer device is similar to other state-of-the-art plasmonic nonlinear modulators based on active layers that were ~20 times thicker[3]. We note that in both optical and SPP pumped measurements, the maximum pump powers we used were conservatively chosen to avoid sample damage. Since the DT signals are linear in pump intensity up to the highest pump powers used (Figs. 3c and 4c), the reported modulation depths should be taken as lower bounds on the achievable modulation depth. In principle, the modulation depth could be further enhanced by using longer TMD layers, stacking several TMD layers separated by hBN, or by decreasing the SPP mode volume by depositing a high dielectric constant material on top of the structure.

To further quantify the performance of our modulator relative to previous works, we consider the minimum energy needed to switch the modulator[3], which is determined by the WSe$_2$ X$^0$ lifetime of ~ 1 ps[27,28] and the SPP pump power of ~40 µW. We obtain an extremely small 40 aJ switching energy per operation which is over an order of magnitude smaller than in previously reported plasmonic modulators[3]. Our architecture could pave the way towards high speed all-plasmonic modulators, amplifiers and transistors based on hybrid 2D material plasmonic structures.

**Materials and Methods:**

**Fabrication:**

The gold waveguide was fabricated on 285 nm $SiO_2$/Si substrates by a two-step electron beam lithography process using an electron beam lithography system (100 kV Ellionix) and a spin coated poly(methyl methacrylate), PMMA, resist. In the first step, 200 nm gold was (electron beam) evaporated onto the substrate using 10 nm titanium sticking layer. In the second lithography step, PMMA was respun and the grating pattern was written and developed. We used an $Ar^+$ milling process to etch the grating couplers into the waveguide. The waveguides are 5 µm x 13 µm. The grating couplers are composed of 5 grooves that are 40 nm deep with a width of 110 nm and period of 570 nm. The bare waveguides were characterized using atomic force microscopy (Supplementary Fig. 4a-d) and optical spectroscopy (Fig. 1c). The waveguide and grating coupler designs were optimized using an FDTD model. Simulations of the bare metallic structure show a maximum transmission of ~4 % at the exciton resonance (Supplementary Fig. 5).

**Figures:**

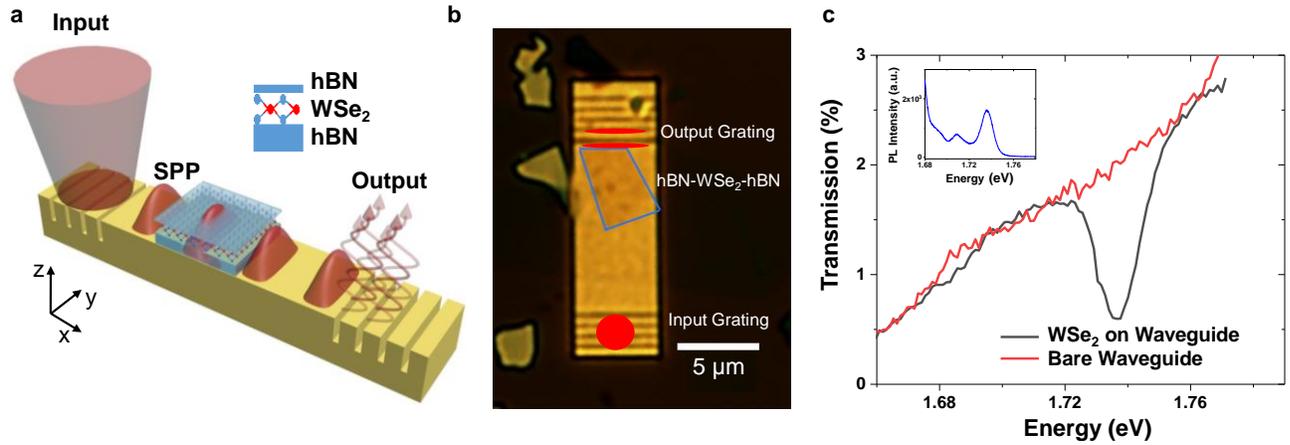

**Fig. 1.** Plasmonic modulator device and linear response. **a** Depiction of the 2D material plasmonic device. SPPs are launched at the input of the device by focusing a free space laser onto an input coupler. The SPPs propagate through the waveguide where they can interact with excitons in the active $WSe_2$ layer, encapsulated in hBN. The SPPs are coupled back to free space photons by an output grating coupler. **b** Optical image of the main device used in this experiment. The red dot shows where the grating was illuminated with the input laser, and the red slits show the typical far field output profile. The blue box shows the location of the active hBN-$WSe_2$-hBN heterostructure. **c** Transmission data of the hBN-$WSe_2$-hBN plasmonic device (black). The normalized transmission of the bare waveguide is shown (red). The inset shows the far-field PL spectrum of the device when excited with a 532 nm (2.33 eV) laser. The sample temperature is 4.5 K.

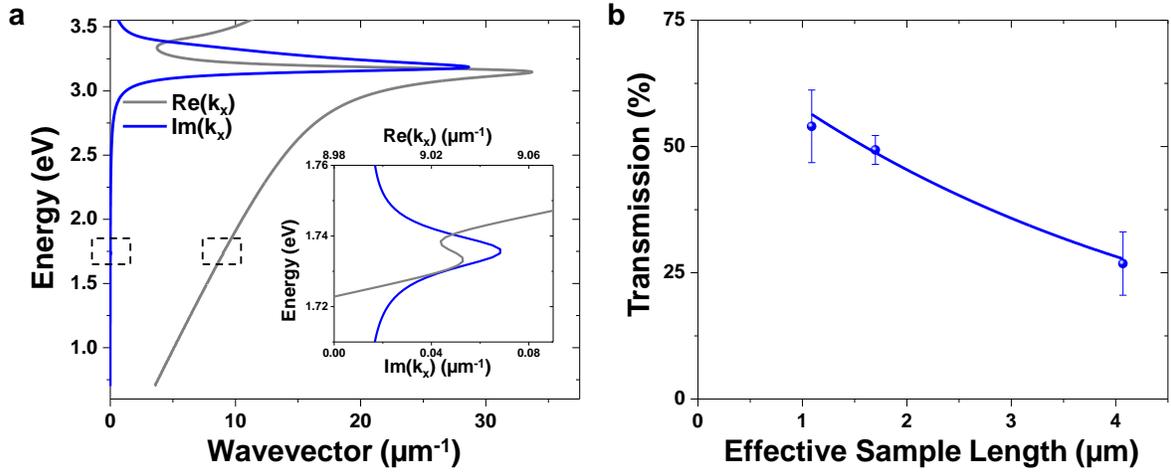

**Fig. 2.** Linear response of exciton surface plasmon polaritons (E-SPPs). **a** Calculated real (grey) and imaginary (blue) parts of the dispersion relation for exciton surface plasmon polaritons, E-SPPs. The inset shows the dispersion near the exciton resonance (1.737 eV), whose location is depicted by dashed boxes on the main figure. **b** Measured transmission at the exciton wavelength as a function of the effective $WSe_2$ sample length on the waveguide.

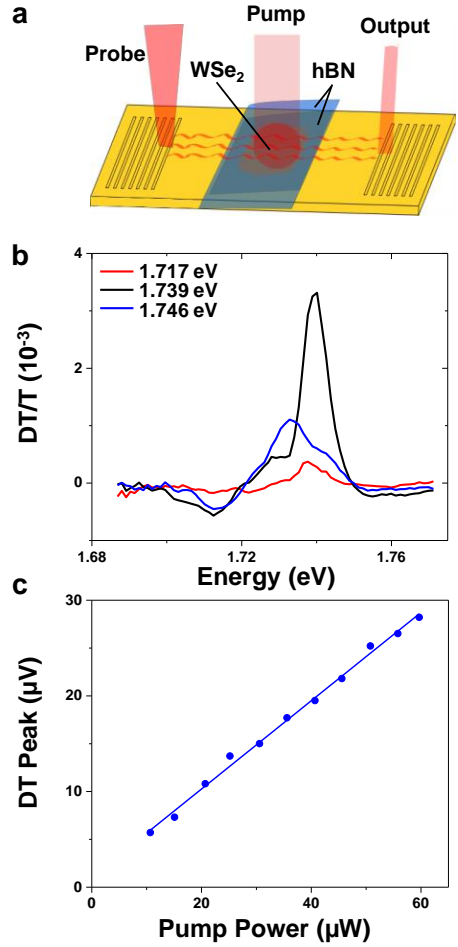

**Fig. 3.** Optical control of SPP propagation. **a** Depiction of the photon-SPP pump-probe measurements. SPPs propagating through the device interact with excitons in the WSe$_2$ layer. A free space laser illuminating the WSe$_2$ controls the SPP propagation by saturating the exciton absorption **b** DT/T measurements for pump photon energies. **c** Pump power dependence of the DT signal near the peak of the exciton response (1.739 eV pump, 1.743 nm probe).

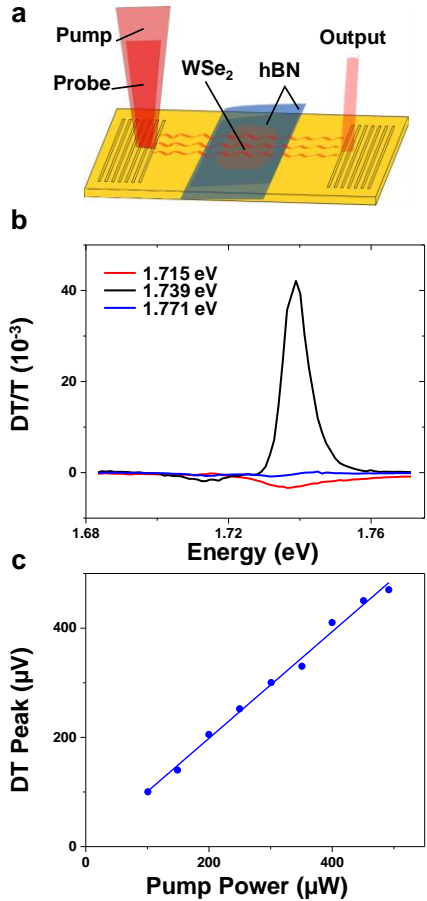

**Fig. 4.** SPP control of SPP propagation. **a** Depiction of the SPP-SPP pump-probe measurements. Pump and probe SPPs are launched at the input grating coupler. SPPs propagating through the device interact with excitons in the WSe$_2$ layer. In this plasmonic modulator configuration, the pump SPP saturates the exciton absorption, resulting in an increase in probe SPP transmission. **b** SPP pump energy dependence of the DT/T spectrum. **c** Pump power dependence of the DT signal near the peak of the exciton response (pump 1.739 eV, probe 1.741 eV).

**Acknowledgments:**

**General:** The authors acknowledge useful discussions with Javier García de Abajo, Robert Norwood, Josh Hendrickson and Ricky Gibson. We thank Ty Newhouse-Illige and Weigang Wang for performing the $Ar^+$ milling process.

**Funding:** This work is mainly supported by the AFOSR-YIP award No. FA9550-17-1-0215. MRK and DGM acknowledge support from the Gordon and Betty Moore Foundation's EPiQS Initiative through grant GBMF4416. KW and TT acknowledge support from the Elemental Strategy Initiative conducted by the MEXT, Japan and the CREST (JPMJCR15F3), JST. JRS acknowledges support from The Science Foundation of Arizona, Bisgrove Scholars Program (Grant No. BSP 0821-17) and AFOSR (Grant No. FA9550-18-1-0049). BJL acknowledges support from the National Science Foundation under Grant No. EECS-1607911. AFM images and data were collected in the W.M. Keck Center for Nano-Scale Imaging in the Department of Chemistry and Biochemistry at the University of Arizona using equipment supported by the National Science Foundation under Grant No. 1337371. The authors would also like to thank Park Systems for their long term loan of the NX20 AFM instrument.

**Author Contributions**:

JRS conceived and supervised the project. MK and BHB fabricated the devices and performed the experiments, assisted by AA and MM. MK, BHB and JRS analyzed the data with input from BJL. MRK and DGM provided and characterized the bulk $WSe_2$ crystals. TT and KW provided hBN crystals. RB developed and evaluated the coupled exciton-SPP (E-SPP) theory. MK, BHB, RB, and JRS wrote the paper. All authors discussed the results.

**Competing Interests:**

The authors declare no competing interests.